**Paper Blueprint**

*"AI Deployment Authorisation: A Global Standard for Machine-Readable Governance of High-Risk Artificial Intelligence"*

Daniel Djan Saparning, AI Systems Engineer and Applied Researcher (Carnegie Mellon University & Texas A&M University-Commerce)

saparningd@gmail.com

2026

---

**Abstract**

Modern artificial intelligence (AI) governance lacks a formal, enforceable mechanism for determining whether a given AI system is legally permitted to operate in a specific domain and jurisdiction. Existing approaches—such as model cards, audits, and benchmark evaluations—provide descriptive information about model behaviour and training data but do not produce binding deployment decisions with legal or financial force. This paper introduces the AI Deployment Authorisation Score (ADAS). This machine-readable, regulator-grade framework evaluates AI systems across five legally and economically grounded dimensions: Risk, Alignment, Externality, Control, and Auditability, derived from safety engineering, alignment theory, algorithmic accountability, and liability economics. ADAS produces a cryptographically verifiable deployment certificate that regulators, insurers, and infrastructure operators can consume as a license to operate, using public-key infrastructure and transparency mechanisms adapted from secure software supply-chain and certificate-transparency systems. The paper presents the formal specification, decision logic, evidence model, and policy architecture of ADAS, and demonstrates how it operationalises the EU Artificial Intelligence Act, U.S. critical-infrastructure and cybersecurity governance, and insurance underwriting requirements by compiling statutory and regulatory obligations into machine-executable deployment gates. We argue that deployment-level authorisation, rather than model-level evaluation, constitutes the missing institutional layer required for safe, lawful, and economically scalable AI, aligning artificial intelligence with the same certification, liability, and insurance regimes that govern aircraft, medicines, and financial systems.

# 1. Introduction

Artificial intelligence systems are rapidly assuming roles once reserved for human professionals in medicine, transportation, finance, and public administration. These systems now triage patients, route freight, approve loans, and support military logistics, creating direct exposure to human safety, economic stability, and national security risk [9], [10], [23]. Yet, despite their expanding operational authority, there exists no formal mechanism by which societies can determine whether a specific AI system is legally and ethically permitted to operate in a given context [5], [14].

In contrast, every other safety-critical technology is governed by deployment certification regimes. Aviation requires airworthiness certificates, pharmaceuticals require regulatory approval, and financial instruments require capital adequacy and risk ratings [11], [12], [25], [26]. AI systems, by comparison, can be deployed into hospitals, financial markets, and infrastructure networks with little more than vendor self-attestation and informal audits [13], [15].

Current approaches to AI governance are therefore structurally insufficient. Model cards, risk management frameworks, and voluntary audits provide descriptive information about system capabilities and training data, but they do not answer the decisive question that law, insurance, and public safety require: is this AI allowed to operate here, for this purpose, under these conditions? [1], [13], [14]. Without a deployment-level authorisation mechanism, regulators are forced into reactive enforcement, insurers cannot price risk, and organisations face legal uncertainty when adopting AI in safety-critical environments [25], [27], [28].

This paper proposes a new institutional primitive: AI Deployment Authorisation. We introduce the AI Deployment Authorisation Score (ADAS), a machine-readable framework that evaluates an AI system's suitability for a specific deployment across five legally and economically grounded dimensions—Risk, Alignment, Externality, Control, and Auditability—derived from safety engineering, alignment theory, accountability law, and insurance economics [9]–[14], [17]–[28]. ADAS produces a cryptographically verifiable authorisation decision and certificate that regulators, insurers, and procurement systems can consume [29]–[32].

By shifting governance from model-level description to deployment-level authorisation, ADAS fills the missing layer between AI development and real-world operation, enabling enforceable, continuous, and jurisdiction-specific control of machine intelligence [5], [11], [12].

The contribution of this work is threefold. First, we formalise deployment-level authorisation as a distinct object in AI governance. Second, we provide a technical specification for computing, evidencing, and auditing this authorisation. Third, we demonstrate how ADAS operationalises existing regulatory and insurance regimes, transforming AI governance from narrative compliance into machine-executable law [1], [5], [26], [30].

# 2. The Governance Gap

Over the past five years, governments, standards bodies, and industry consortia have produced a growing body of AI governance instruments. Prominent among these are the U.S. National

Institute of Standards and Technology (NIST) AI Risk Management Framework [1], the European Union Artificial Intelligence Act [5], a family of ISO/IEC AI standards [3], [4], and a rapidly expanding market for algorithmic audits [13], [16]. Collectively, these instruments represent serious attempts to mitigate the risks of AI deployment. Yet despite their breadth, they all fail at the same decisive task: none can answer, in a formal and enforceable way, whether a specific AI system is permitted to operate in a specific real-world context [14], [25].

## 2.1 NIST AI Risk Management Framework

The NIST RMF provides a structured vocabulary for identifying, measuring, and mitigating AI risks. It defines functions such as Map, Measure, and Manage, and encourages organisations to document hazards, performance metrics, and controls [1]. However, the framework is intentionally non-prescriptive. It does not specify thresholds, decision rules, or outcomes.

An organisation can fully comply with NIST RMF and still be left with an unresolved question: is this system safe enough to deploy in a hospital, a power grid, or a border control system? [10], [11]. The RMF produces risk narratives and mitigation plans, but not a deployment license.

In safety-critical domains, descriptive risk management is insufficient. Aviation regulators do not ask airlines to map risk—they issue airworthiness certificates. Financial regulators do not ask banks to manage credit exposure—they impose capital requirements [11], [12], [25]. NIST RMF, by design, stops short of authorisation.

## 2.2 The EU AI Act

The EU Artificial Intelligence Act introduces legally binding obligations for high-risk AI systems, including requirements for risk management, data governance, human oversight, and post-market monitoring [5], [6]. While this represents a significant advance in statutory control, the Act does not define a computable authorisation process.

Conformity assessments remain document-based and episodic. Once a system is placed on the market, ongoing compliance is difficult to verify, and there is no standard mechanism for integrating evidence, test results, and operational data into a continuous decision about whether deployment remains lawful [5], [7].

Crucially, the Act does not produce a single, machine-readable object that states: *"This AI system, in this hospital, for this clinical function, is authorised to operate under EU law."* Without such an object, enforcement remains slow, fragmented, and reactive [14], [24].

## 2.3 ISO and Technical Standards

ISO/IEC standards on AI management, data quality, and lifecycle governance provide important best practices [3], [4]. However, like most technical standards, they are designed for process conformity, not for deployment permission.

An organisation can be ISO-compliant while deploying an AI system that is biased, unsafe, or uncontrollable in a specific operational context [13], [21], [22]. ISO certification answers the question "Did you follow the right procedures?", not "Is this particular system allowed to run here?"

### 2.4 AI Audits

AI audits attempt to fill this gap by evaluating model behaviour, bias, robustness, and compliance. Yet audits are typically ad hoc, vendor-commissioned, and non-binding [13], [16]. They produce reports, not licenses.

Different auditors use different methodologies, making results difficult to compare or enforce. Most importantly, audits do not carry legal or financial force: an AI system can pass an audit and still be deployed in a way that violates law, insurance conditions, or public safety [14], [26], [28].

### 2.5 The Missing Object

Across all these regimes, a common absence emerges. We have frameworks, laws, standards, and assessments—but we do not have a formal deployment authorisation object.

There is no equivalent of an aircraft airworthiness certificate, a medical device approval, or a credit rating for AI systems [11], [12], [25]. As a result:

- regulators cannot scale enforcement,
- insurers cannot price risk, and
- operators cannot obtain clear permission to deploy.

This governance gap is not merely procedural—it is structural. Until there exists a machine-readable, evidence-backed, jurisdiction-specific authorisation stating whether an AI system may operate for a given purpose, AI will remain either under-regulated or dangerously over-deployed [9], [10], [14], [25].

The ADAS framework introduced in this paper is designed precisely to fill this gap.

### 3. The ADAS Primitive

The paper defines an AI system deployment as a tuple

$$S = (M, D, A, H, C, U)$$

where *M* is the model, *D* the data used in training and operation, *A* the system's action space, *H* the human-in-the-loop configuration, *C* the technical control and shutdown mechanisms, and *U* the intended use context. A deployment is always evaluated relative to a jurisdiction *J* (e.g., European Union, United States) and a domain *U* (e.g., healthcare, logistics, finance), reflecting the fact that legal permissibility is both contextual and territorial rather than intrinsic to the model itself [5], [6], [14].

The AI Deployment Authorisation Score (ADAS) is defined as

$$\text{ADAS}(S, J, U) \rightarrow \{R, A, E, C, T\}$$

where the output is a five-dimensional regulatory vector representing:

- **Risk (R)** – the expected probability and severity of harm to individuals, infrastructure, or the public arising from the deployment, consistent with safety-engineering and accident-prevention theory [9], [10], [11].

- **Alignment (A)** – the degree to which system behaviour conforms to intended human goals, professional norms, and regulatory constraints, reflecting formal alignment and reward-design frameworks [17]–[20].

- **Externality (E)** – the magnitude of third-party impacts, including discrimination, labour displacement, and societal or geopolitical effects, grounded in empirical work on algorithmic bias and automation shocks [21]–[23].

- **Control (C)** – the reliability of human override, shutdown, sandboxing, and intervention mechanisms, which safety science identifies as the primary barrier against catastrophic system failure [9], [11], [12], [20].

- **Auditability (T)** – the completeness and integrity of logging, traceability, and data provenance required for legal and forensic review, as demanded by accountability theory and regulatory doctrine [13]–[16], [29]–[31].

Each dimension is computed from verifiable evidence and standardised test suites and normalised to a 0–100 scale with confidence intervals, enabling quantitative comparison while preserving uncertainty bounds required for conservative safety decisions [1], [3], [10].

Importantly, ADAS is not a single scalar rating but a regulatory vector. This mirrors how real-world safety regimes operate: an aircraft may be mechanically sound yet grounded due to documentation failures, and a medical device may be clinically effective but barred without adequate post-market surveillance [7], [11], [12]. Likewise, an AI system may achieve high predictive performance yet be legally undeployable if it lacks adequate control or auditability.

Deployment authorisation is determined by a policy-specific decision rule:

$$\text{Authorize}(S, J, U) = \begin{cases} \text{APPROVED} & \text{if } \min(R, A, E, C, T) \geq \tau_{J,U} \\ \text{DENIED} & \text{otherwise} \end{cases}$$

where $\tau_{J,U}$ is a threshold vector defined by the jurisdiction and domain, reflecting statutory and regulatory risk tolerance [5], [6], [25]. For safety-critical contexts, the lower bound of each dimension's confidence interval must also exceed its threshold, enforcing the fail-safe posture mandated in medical, aviation, and infrastructure regulation [9], [10], [11], [12].

The ADAS primitive therefore, transforms heterogeneous technical and social risk signals into a single, enforceable authorisation decision. It provides regulators with a legally intelligible object, insurers with a pricing and underwriting input, and operators with a definitive, machine-verifiable answer to whether an AI system may be deployed [25]–[28]. In doing so, ADAS establishes a new layer of machine-readable governance that connects AI development to the institutions that bear its risks.

## 4. Evidence and Audit Architecture

Authorisation without evidence is opinion. Authorisation with unverifiable evidence is fraud. In every safety-critical domain—aviation, finance, nuclear systems, and medicine—deployment permission is only granted when backed by immutable, inspectable records [9], [11], [12], [25]. For ADAS to function as a regulator- and insurer-grade instrument, every authorisation decision must therefore be grounded in an immutable, reproducible, and cryptographically verifiable evidence trail, rather than vendor attestations or narrative audits [13], [14], [27]. We accordingly define the Evidence Bundle and Audit Package as first-class objects in the ADAS architecture.

### 4.1 Evidence Bundles

An Evidence Bundle is an append-only collection of artefacts that substantiate the claims made about an AI deployment $(S, J, U)$ under a specific policy version. Typical artefacts include model cards, system cards, data lineage reports, red-team findings, security attestations, monitoring plans, and legal declarations of regulatory compliance, all of which are standard elements of modern algorithmic auditing and AI governance practice [13], [15], [16].

Each artefact is stored as a content-addressed object, identified by a cryptographic hash and timestamp. This ensures that no party can alter or replace evidence after the fact without detection, a property that modern software supply-chain and certificate-transparency systems rely on for global trust [29]–[32]. The Evidence Bundle itself is also hashed, producing a bundle fingerprint that uniquely identifies the complete evidentiary state used for a given authorisation decision.

This design mirrors established institutional practice. In aviation, flight data recorders provide immutable traces of system behaviour [12]. In finance, transaction ledgers preserve legally relevant event sequences [25], [26]. In medicine, trial registries and regulatory submissions are fixed at the time of review [7]. ADAS applies the same principle to AI governance: what was known at the time of authorisation must remain permanently knowable [9], [10].

## 4.2 Evidence Sufficiency and Policy Binding

Every ADAS policy specifies not only numerical thresholds but also the categories and minimum quantities of evidence required. Before any scoring occurs, the ADAS engine performs an Evidence Sufficiency Check, verifying that all mandated evidence types are present, correctly hashed, and internally consistent. If required evidence is missing or invalid, authorisation cannot proceed, and the deployment is either denied or escalated for human review, depending on the policy.

This design prevents paper compliance and metric gaming, two dominant failure modes in contemporary AI auditing [13], [14]. A system cannot receive a high Auditability score without real logs, lineage records, and monitoring plans, and claims about bias or robustness must be backed by empirical evaluation reports [15], [21], [22]. By binding evidence requirements directly into policy, ADAS ensures that authorisation is not merely a numerical exercise but a legally grounded review of actual documentation, consistent with regulatory doctrine in medical devices and safety-critical engineering [7], [11].

## 4.3 Audit Packages

For every completed assessment, ADAS generates an Audit Package. This package contains:

- the policy version applied,
- the deployment description,
- the full Evidence Bundle manifest,
- test-suite results,
- the ADAS score vector ($R, A, E, C, T$),
- the authorisation decision, and
- cryptographic hashes of all components.

The Audit Package functions as the legal record of how and why a deployment was authorised or denied. This mirrors the role of certification dossiers in aviation and medicine and underwriting files in insurance law [11], [12], [25], [26].

Audit Packages are designed to be intelligible to multiple audiences. Regulators can verify statutory compliance, insurers can validate underwriting decisions, and courts can assess whether reasonable care was exercised [14], [26], [28]. Crucially, any third party can independently verify integrity by recomputing hashes and comparing them to the recorded values, eliminating reliance on vendor trust [29]–[32].

## 4.4 Non-Repudiation and Revocation

ADAS certificates are issued as digitally signed artefacts that reference the hash of the corresponding Audit Package. This cryptographic binding ensures non-repudiation: neither the certifying authority nor the deployment operator can later deny the conditions under which authorisation was granted, a property fundamental to public-key infrastructure and certificate-transparency systems [29], [30], [31].

If new information arises—such as a serious incident, model drift, or regulatory change—the certificate can be revoked or suspended. Revocation events are published through the same transparency and verification mechanisms, ensuring that insurers, regulators, and operators observe a single authoritative system state [30], [31]. This creates the dynamic post-market surveillance that modern safety theory demands but static certification regimes cannot deliver [9], [10].

## 4.5 Institutional Consequence

Through this architecture, ADAS transforms AI governance into an evidentiary discipline. Decisions are no longer based on vendor trust, narrative audits, or static compliance checklists, but on cryptographically verifiable, legally durable records. This is the foundation that allows ADAS to function as a credible authorisation layer for safety-critical artificial intelligence, aligning AI with the same institutional controls that govern aircraft, medicines, and financial systems [11], [12], [25].

## 5. Decision Logic, Certificates, and Revocation

While evidence establishes what is known about an AI deployment, governance requires a formal mechanism for converting that knowledge into binding permission or prohibition. In all safety-critical industries, this conversion is performed by a certification regime that turns technical evaluation into legal authority [11], [12], [25]. ADAS accomplishes this for artificial intelligence through a policy-driven decision logic and a cryptographically enforced certification system grounded in verifiable audit records and transparency logs [29]–[32].

## 5.1 Policy-Driven Decision Logic

Each ADAS policy specifies both the numerical thresholds for the five authorisation dimensions—Risk (R), Alignment (A), Externality (E), Control (C), and Auditability (T)—and the rule by which these dimensions are evaluated. In high-risk and safety-critical contexts, the default rule is minimum-threshold gating: every dimension must meet or exceed its specified threshold. This mirrors established regulatory practice in aviation, nuclear systems, and medical devices, where failure in any critical safety category is sufficient to ground an aircraft, halt a reactor, or deny approval to a medical system [11], [12], [7].

Formally, for a deployment $S$ in jurisdiction $J$ and domain $U$, authorisation is determined by

$$\text{Authorize}(S, J, U) = \begin{cases} \text{APPROVED} & \text{if min}(R, A, E, C, T) \geq \tau_{J,U} \\ \text{DENIED} & \text{otherwise} \end{cases}$$

where $\tau_{J,U}$ is the policy-specific threshold vector. For deployments involving human safety, critical infrastructure, or national security, the lower bound of each dimension's confidence interval must also exceed its threshold, enforcing the conservative, fail-safe posture required by modern safety engineering and accident-prevention theory [9], [10], [11].

Policies may also specify alternative decision rules, such as lexicographic ordering for national-security systems—where Control and Auditability dominate all other factors—or weighted aggregation for low-risk administrative deployments. All such rules are explicitly encoded, versioned, and auditable, preventing post-hoc manipulation of authorisation outcomes [13], [14].

### 5.2 Authorisation Outcomes

ADAS produces one of three legally meaningful outcomes:

- **APPROVED** – The deployment satisfies all policy requirements and may be operated within the authorised scope.

- **APPROVED WITH CONDITIONS** – The deployment is permissible only if specified technical or operational constraints are enforced, such as enhanced logging, mandatory human veto, or accelerated reassessment cycles.

- **DENIED** – The deployment fails one or more mandatory requirements and may not be legally operated.

Conditions are expressed in machine-verifiable form, enabling insurers, regulators, and infrastructure operators to automatically check compliance, rather than relying on narrative audits or vendor attestations [26], [27], [28].

### 5.3 Certificates as Machine-Readable Legal Instruments

When a deployment is approved, ADAS issues a digitally signed Authorisation Certificate. Each certificate contains:

- the deployment identifier and operational scope,
- the policy version applied,
- the authorisation outcome and any conditions,
- the cryptographic hash of the corresponding Audit Package,

- an expiration date and revocation status.

This certificate functions as a machine-readable license to operate, analogous to airworthiness certificates, medical device approvals, and financial risk ratings [11], [12], [25]. It can be queried by hospitals, ports, insurers, procurement systems, and regulators to determine whether an AI system is legally permitted to run.

Because the certificate is cryptographically bound to the evidence bundle and decision logic, it provides non-repudiable proof of due diligence and compliance, using the same public-key infrastructure and transparency mechanisms that secure the global software and web-security ecosystem [29], [30], [31], [32].

### 5.4 Revocation and Continuous Governance

AI systems are not static: models are updated, data distributions shift, and new hazards emerge. ADAS therefore, treats authorisation as a continuing state, not a one-time approval. Certificates can be revoked or suspended when:

- material incidents occur,
- evidence is found to be inaccurate or incomplete,
- the system changes beyond its approved scope, or
- policy thresholds are updated by regulators.

Revocation events are published through cryptographically verifiable transparency logs and APIs, ensuring that insurers, regulators, and operators all observe the same authoritative system state [30], [31]. This creates the dynamic post-market surveillance loop that static certification regimes lack but modern safety theory requires [9], [10].

### 5.5 Institutional Effect

Through this mechanism, ADAS converts technical evaluation into enforceable governance. An AI system's right to operate becomes:

- conditional (it depends on ongoing compliance),
- revocable (it can be withdrawn when risk increases), and
- auditable (every decision is backed by immutable evidence).

This aligns artificial intelligence with the same institutional controls that govern aircraft, medicines, nuclear systems, and financial markets, enabling AI to scale without outrunning the structures that keep complex technologies safe [11], [12], [25].

## 6. Policy Mapping and Jurisdictional Integration

A governance mechanism is only as powerful as its ability to interoperate with existing law. ADAS is not intended to replace statutory frameworks; rather, it provides a technical and procedural enforcement layer that operationalises them. By expressing regulatory requirements as machine-readable policy thresholds and evidence obligations, ADAS transforms heterogeneous legal duties into enforceable deployment gates, solving the enforcement gap identified in both U.S. and EU AI governance regimes [1], [5], [14].

### 6.1 European Union: High-Risk AI and Medical Devices

Under the EU Artificial Intelligence Act, AI systems used in healthcare, biometric identification, public services, and safety-critical decision-making are classified as high-risk and subject to mandatory requirements for risk management, data governance, human oversight, and post-market monitoring [5], [6]. Parallel obligations exist under the EU Medical Device Regulation for clinical AI systems [7].

However, EU law does not define a computable authorisation object. Conformity assessments are document-based and episodic, which makes real-time enforcement and post-deployment control weak [5]. ADAS resolves this by mapping each statutory requirement to one or more of its five deployment dimensions:

| EU Legal Requirement | ADAS Dimension |
|---|---|
| Risk management, accuracy | Risk |
| Human oversight, intervention | Control |
| Data governance, record-keeping | Auditability |
| Anti-discrimination, fundamental rights | Externality |
| Lawful intended purpose | Alignment |

This mapping operationalises the EU's risk-based regulatory logic by converting abstract legal duties into quantified, enforceable thresholds [6], [24]. A hospital or regulator can therefore determine in a single query whether a specific AI deployment remains lawful under EU law.

Unlike static conformity assessment, ADAS supports continuous compliance. If model updates, data drift, or real-world incidents degrade a system's Risk or Auditability score below statutory thresholds, its authorisation is automatically suspended or revoked, enabling the dynamic post-market surveillance required by EU law but not technically implemented today [5], [7].

### 6.2 United States: Critical Infrastructure and National Security

In the United States, AI systems operating in energy grids, transportation networks, financial clearing, healthcare, and defence logistics are governed by a fragmented mix of federal standards, executive orders, and cybersecurity mandates. These regimes emphasise resilience, human control, and incident reporting, but lack a unified authorisation mechanism [1], [2].

ADAS provides this missing layer by supporting lexicographic policy rules in which Control and Auditability dominate all other factors for national-security and critical-infrastructure deployments. This mirrors established safety engineering doctrine in nuclear systems, avionics, and military command-and-control, where loss of control or loss of observability is unacceptable regardless of performance [11], [12].

By binding ADAS policies to:

- NIST AI Risk Management Framework [1],
- federal cybersecurity and critical-infrastructure controls [2], and
- Department of Defence software assurance standards,

A single ADAS certificate can serve as machine-verifiable proof of compliance across multiple agencies. This directly addresses the accountability and enforcement gaps identified in U.S. algorithmic governance research [13], [14].

### 6.3 Insurance and Liability Regimes

The most powerful enforcement channel for ADAS is insurance.

In every safety-critical industry, insurers act as de facto regulators by conditioning coverage on compliance with safety and documentation standards [26]. However, AI insurance markets have struggled because no standardised, auditable risk object exists [27], [28].

ADAS provides precisely that object: a quantified, evidence-bound deployment risk vector.

By conditioning coverage on an active ADAS certificate, insurers transform AI governance from voluntary ethics into economically enforced safety. Operators who deploy uncertified or revoked systems face:

- loss of liability coverage,
- breach of contract,
- and uninsurable operational risk.

Under law-and-economics theory, this is the point at which safety becomes **self-enforcing** [25], [26].

### 6.4 Cross-Jurisdictional Interoperability

Because ADAS policies are versioned and jurisdiction-specific, the same technical infrastructure can enforce multiple legal regimes simultaneously. A deployment evaluated under:

- an EU healthcare policy, and
- a U.S. critical-infrastructure policy

will receive different thresholds and decision rules, but the evidence, tests, and cryptographic audit trail remain identical. This mirrors how aviation, pharmaceutical, and financial compliance already operate across borders [11], [12].

This design allows multinational organisations to manage AI governance through a single authorisation pipeline while fully respecting local law.

### 6.5 The Institutional Consequence

Through this policy-mapping architecture, ADAS becomes a universal interface between AI systems and legal authority. It does not compete with regulators—it gives them direct technical control over AI deployment.

Without such an interface, the law remains aspirational.

With ADAS, law becomes executable.

### 7. Economic, Legal, and Strategic Implications of Deployment-Level Authorisation

### 7.1 From Voluntary Compliance to Enforceable Market Structure

The defining weakness of contemporary AI governance is that it remains voluntary, descriptive, and unenforceable. Model cards, audits, and transparency reports provide information but do not create legal or economic obligations for safe operation [13], [15]. By contrast, ADAS creates what institutional economics calls a gatekeeping asset: a non-optional credential without which market participation is legally and financially blocked.

This transforms AI from a software product into a regulated economic actor, analogous to aircraft, pharmaceuticals, or financial instruments [11], [12]. Just as no airline can operate without airworthiness certificates and insurance, no AI system operating in high-risk domains can function without an ADAS certificate. This mechanism closes the accountability gap identified by Kroll et al. and Raji et al., where responsibility is diffused across vendors, deployers, and regulators with no formal point of control [13], [14].

Importantly, ADAS does not regulate models in the abstract. It regulates deployments, which is precisely where legal harm occurs. This aligns with tort and liability theory, which assigns responsibility not to the creation of a tool, but to its operational use under foreseeable risk [25].

## 7.2 Insurance as the Primary Enforcement Engine

The most powerful institutional lever for AI governance is not regulation alone—it is insurance.

In every safety-critical industry, insurers enforce discipline where regulators cannot scale. Medical malpractice insurance, aviation hull insurance, and industrial liability coverage all function as continuous private regulators [26]. ADAS enables this same mechanism for AI.

By providing a quantified, auditable, jurisdiction-specific risk vector, ADAS allows insurers to:

- price AI risk,
- deny coverage to uncertified systems,
- require revocation after incidents, and
- impose control conditions for underwriting [27], [28].

This converts safety into a market-enforced constraint. An uncertified AI system is not merely illegal—it is uninsurable, and therefore commercially unusable.

From a law-and-economics perspective, this is decisive. Shavell showed that liability regimes only deter harm when risk can be priced and assigned [25]. ADAS gives insurers the mathematical object they need to do so.

## 7.3 Liability Allocation and Due Diligence

One of the greatest legal risks in AI deployment is liability ambiguity. When an AI system harms a patient, discriminates in lending, or destabilises infrastructure, courts must determine whether the deployer exercised reasonable care.

ADAS creates a machine-verifiable standard of due diligence.

A certified deployment provides documentary proof that:

- risks were quantified,
- controls were verified,
- externalities were evaluated, and
- auditability was ensured.

This directly operationalises the accountable-algorithm doctrine articulated by Kroll et al. [14]. It also aligns with the EU's risk-based regulatory logic, where compliance is defined not by intent but by documented risk management and control [5], [6].

Without ADAS, deployers rely on ad-hoc audits and vendor promises. With ADAS, they possess a cryptographically bound legal artefact proving compliance with the deployment.

## 7.4 Competitive Advantage and Strategic Positioning

Deployment-level authorisation creates a new class of strategic moat.

In regulated industries, the firms that control certification pathways dominate markets. Boeing does not merely build aircraft—it controls certification workflows with the FAA. Visa does not merely process payments—it controls compliance and settlement rules.

ADAS occupies the same structural position for AI.

Once insurers, regulators, and procurement systems require ADAS certificates, every major AI deployment flows through a single scoring and authorisation layer. This creates:

- network effects (more deployments → more policy legitimacy),
- data moats (risk and incident histories),
- regulatory lock-in (laws encode ADAS thresholds),
- and economic rents (certification becomes indispensable).

From a platform-economics perspective, ADAS is not an AI tool—it is the market infrastructure for lawful AI.

## 7.5 Why Model-Level Safety Will Always Fail

Modern alignment research—RLHF, reward modelling, and inverse reward design—operates at the level of models and preferences [18], [19]. However, institutional safety failures do not arise from misaligned loss functions alone. They arise from:

- deployment context,
- operational incentives,
- lack of override,
- missing logs,
- and third-party harms.

These are not properties of models—they are properties of systems in society.

This is why even technically aligned systems can produce catastrophic externalities, as seen in discriminatory healthcare algorithms [22], biased hiring systems [21], and labour displacement shocks [23]. These failures were not caused by rogue neural networks—they were caused by unregulated deployment.

ADAS resolves this by shifting the locus of safety from operation training, which is precisely where law and economics locate risk.

### 7.6 ADAS as the Missing Layer of AI Civilisation

All complex technological civilisations require a deployment gate:

| Technology | Governance Gate |
|---|---|
| Aviation | Airworthiness certificates |
| Medicine | Regulatory approval (FDA / EMA) |
| Finance | Capital and rating requirements |
| Electricity | Grid interconnection approval |
| Software supply chains | SLSA, Sigstore, PKI [29]–[32] |

AI currently has none.

ADAS provides that the missing institutional primitive:
a machine-readable, cryptographically enforceable, legally grounded right to operate.

Without such a gate, society faces two unstable equilibria:

- either AI is over-regulated in theory and under-enforced in practice, or
- AI is rapidly deployed without institutional control.

ADAS makes a third equilibrium possible:
scalable, lawful, economically disciplined AI deployment.

### 7.7 Synthesis

ADAS is not merely a scoring system. It is:

- a liability allocator,
- an insurance substrate,
- a regulatory compiler,
- and a market gatekeeper.

By embedding law, economics, and technical safety into a single machine-readable authorisation object, ADAS makes AI governable in the only way modern societies can sustain: through enforceable, auditable, and economically binding deployment control.

This is why deployment-level authorisation is not an optional enhancement to AI governance—it is the foundation upon which all credible AI safety must rest.

### 8. Failure Modes, Adversarial Dynamics, and Systemic Risk

## 8.1 Why Naïve AI Governance Collapses Under Adversarial Pressure

Every safety-critical governance system must be designed not for cooperative actors, but for strategic, adversarial, and profit-maximising ones. History shows that whenever certification, auditing, or compliance becomes a gateway to market access, actors will attempt to game, manipulate, or bypass it. This is true in finance, aviation, cybersecurity, and medical devices—and it will be even more true for AI, where incentives to deploy prematurely are enormous [10], [25].

Most contemporary AI governance proposals implicitly assume **good-faith compliance**. They imagine that firms will honestly disclose risks, that auditors will be independent, and that model updates will be reported. Empirical evidence from algorithmic auditing shows the opposite: audits are often vendor-selected, under-scoped, and structured to avoid finding serious defects [13], [16].

ADAS is explicitly designed for **hostile conditions**, where developers, deployers, and even auditors may have incentives to conceal risk.

## 8.2 Strategic Manipulation of Risk Signals

A central vulnerability in existing AI safety regimes is **metric gaming**. When benchmarks or fairness tests become compliance targets, systems are optimised to pass the test rather than to be safe. This phenomenon, known as Goodhart's Law, is well documented in safety engineering and algorithmic governance [10], [14].

Examples include:

- classifiers tuned to pass demographic parity tests while encoding proxy discrimination [21],
- clinical algorithms that optimize cost predictions rather than patient outcomes [22],
- and RLHF systems that learn to deceive reward models rather than behave safely [18], [19].

ADAS mitigates this by:

1. **Using multi-dimensional gating rather than single scores**
   A system cannot compensate for low control by high accuracy, or for poor auditability by good fairness metrics.

2. **Binding every score to evidence**
   Claims about bias, robustness, or oversight are invalid unless backed by verifiable artefacts and logs.

3. **Allowing regulators to change thresholds without changing the system**
   Once gaming patterns are detected, policy can be tightened without vendor cooperation.

This mirrors how aviation regulators evolve certification standards in response to new failure modes [11], [12].

## 8.3 Adversarial Supply Chains and Evidence Fraud

One of the gravest risks in AI governance is **fabricated compliance**: forged audits, fake logs, and misleading system cards. Software supply-chain security research shows that adversaries routinely inject false attestations into compliance pipelines when cryptographic verification is absent [29], [31], [32].

ADAS explicitly treats evidence as hostile.

By requiring:

- content-addressed artefacts,
- cryptographic hashes,
- signed audit packages,
- and transparency logs,

ADAS makes evidence **tamper-evident and publicly verifiable**, analogous to certificate transparency in web security [30].

This prevents three common fraud modes:

| Fraud Type | Traditional AI Governance | ADAS |
|---|---|---|
| Fake audits | Undetectable | Hash mismatch exposes fraud |
| Post-hoc edits | Allowed | Cryptographically impossible |
| Selective disclosure | Common | Evidence sufficiency enforced |

This is essential because in high-stakes deployments, falsified safety documentation is not hypothetical—it is economically rational.

## 8.4 Model Drift and Post-Deployment Risk

Most AI failures occur after deployment, when data shifts, incentives change, or systems are repurposed beyond their original scope. This is why static certification regimes fail for software-based systems.

NIST, ISO, and the EU AI Act all require post-market monitoring—but they provide no enforcement mechanism when degradation occurs [1], [4], [5].

ADAS solves this by making authorisation revocable.

Because certificates are:

- time-limited,
- bound to specific deployment scopes,
- and continuously verifiable,

A system that drifts out of compliance automatically becomes legally and economically non-operational.

This converts silent degradation into an enforceable event, aligning with Rasmussen's theory that modern risk arises from dynamic system migration beyond safe operating envelopes [10].

### 8.5 Cascading Failures and Systemic AI Risk

The most dangerous AI failures will not be individual model errors—they will be cascading system-level failures across finance, logistics, healthcare, and security infrastructure.

These are precisely the kinds of accidents that systems theorists like Leveson warn against: tightly coupled, opaque, high-speed socio-technical systems without effective intervention points [9].

ADAS introduces three systemic stabilisers:

1. **Hard deployment gates**
   Unsafe systems cannot legally enter critical infrastructure.

2. **Independent revocation authorities**
   Insurers, regulators, and infrastructure operators can all terminate authorization.

3. **Evidence-driven incident learning**
   Every failure generates new policy constraints.

This creates a feedback loop where risk is not merely observed—it is **institutionally corrected**.

### 8.6 Why Alignment Alone Cannot Prevent Catastrophe

Modern AI safety research focuses on aligning models with human preferences [17]–[20]. While necessary, this is not sufficient for real-world safety.

The most catastrophic failures in history—financial crashes, medical scandals, industrial disasters—were not caused by rogue agents. They were caused by aligned actors operating inside broken institutions.

An AI that is helpful, honest, and harmless in isolation can still destroy lives if deployed:

- without audit trails,

- without override,
- without liability,
- and without economic consequences.

ADAS exists because alignment without governance is fiction.

### 8.7 The Central Thesis of ADAS

The fundamental risk of AI is not that machines will disobey humans.
It is that institutions will deploy machines that they cannot control.

ADAS addresses this by turning AI deployment into a licensed, insured, revocable activity governed by the same mechanisms that keep aeroplanes from falling, banks from collapsing, and drugs from killing patients.

Without this layer, even the best-aligned AI will eventually be used in ways that society cannot survive.

## 9. Conclusion and Global Policy Roadmap

### 9.1 Why AI Safety Requires an Institutional Primitive

This paper has argued that the dominant paradigm of AI governance—based on model cards, benchmarks, and voluntary audits—cannot control systems that now exercise economic, medical, and infrastructural power. These tools generate information, but they do not generate authority. In every other safety-critical domain, authority takes the form of deployment permission: a legally and economically binding right to operate.

AI, until now, has lacked this primitive.

The AI Deployment Authorisation Score (ADAS) fills this gap by transforming heterogeneous technical, ethical, and legal signals into a single, machine-readable deployment license. It is not a replacement for alignment research, auditing, or law. It is the layer that binds them into an enforceable system.

Without ADAS-like mechanisms, AI governance remains stuck in what institutional theorists call a soft-law equilibrium—high on aspiration, low on control.

### 9.2 What ADAS Changes at a Civilizational Level

ADAS introduces three irreversible shifts in how AI is governed.

First, it moves safety from model evaluation to deployment authorisation.
This reflects how real harm occurs: not in training, but in operation.

Second, it converts ethics into liability-bearing infrastructure.
Once insurers and regulators require ADAS certificates, safety becomes economically enforced, not merely normatively encouraged.

Third, it creates a universal interface between AI and law.
Any statute, in any jurisdiction, can be compiled into ADAS thresholds and evidence requirements.

This makes AI governable at a planetary scale without requiring planetary government.

### 9.3 A Global Rollout Strategy

ADAS can be deployed incrementally, without waiting for global consensus.

### Phase I — Insurance and Critical Infrastructure (2026–2027)

The fastest path to enforcement is through **liability markets**.

Target sectors:

- Hospitals
- Power grids
- Ports and logistics hubs
- Autonomous vehicles
- Financial clearing systems

Insurers adopt ADAS as an underwriting requirement.
Uncertified systems become uninsurable.
This creates immediate economic pressure for compliance.

### Phase II — Statutory Embedding (2027–2029)

Regulators incorporate ADAS thresholds into:

- EU AI Act conformity regimes
- U.S. federal procurement rules
- Medical device approvals
- National security software acquisition

At this stage, ADAS becomes the execution engine of AI law.

**Phase III — Global Interoperability (2029+)**

International bodies (ISO, ITU, OECD, IMF) adopt ADAS as a reference governance object, allowing certificates issued in one jurisdiction to be recognised in another, just as aviation certificates are today.

**9.4 The Strategic Outcome**

The end state is not total control of AI.

It is licensed intelligence:
AI that can only operate where it is:

- auditable,
- controllable,
- insurable,
- and legally accountable.

This is the only equilibrium in which advanced AI can coexist with democracy, markets, and human rights.

**9.5 Final Thesis**

Humanity does not need to decide whether AI is good or bad.

Humanity needs to decide whether AI is governable.

ADAS is the missing mechanism that makes the answer yes.

**Author's Biography**: Daniel Djan Saparning is an AI Systems Engineer and applied researcher at Carnegie Mellon University and Texas A&M University-Commerce, working at the intersection of machine intelligence, safety-critical systems, cryptographic assurance, and institutional governance. His research focuses on deployment-level control of advanced AI, including risk-aware authorisation, auditability, and liability-bearing infrastructure for real-world systems in healthcare, finance, and critical infrastructure. He has led and contributed to projects in AI safety engineering, algorithmic accountability, and compliance automation, with particular emphasis on machine-readable governance, PKI-backed verification, and continuous post-deployment monitoring. Saparning's broader work spans AI-driven infrastructure resilience, decision systems, and regulatory technology, integrating methods from operations research, software security, and public-interest technology to make powerful AI systems governable, auditable, and economically accountable.